\documentclass[letterpaper, 10 pt, conference]{ieeeconf}

\usepackage{graphicx}
\IEEEoverridecommandlockouts                 
\usepackage{amsmath,amssymb,amsfonts}
\usepackage{subfigure}
\usepackage{graphicx}
\graphicspath{{./IMG/}}
\usepackage{epstopdf}
\usepackage{cite}
\usepackage{hyperref}
\usepackage{epstopdf,color}
\usepackage{textcomp}
\usepackage{enumerate}

\usepackage{color,array}
\usepackage{comment}
\usepackage{amsmath}
\usepackage{etoolbox}
\usepackage{algorithm,algorithmic}
\usepackage{mathrsfs}
\AtBeginEnvironment{algorithm2e}{\algoequations}
\AtEndEnvironment{algorithm2e}{\restoreequations}
\newcounter{algosavedequation}
\newcommand{\algoequations}{%
  \setcounter{algosavedequation}{\value{equation}+1}%
  \setcounter{equation}{0}%
  \renewcommand{\theequation}{\arabic{algosavedequation}\alph{equation}}%\thealgocf
}
\newcommand{\restoreequations}{%
  \setcounter{equation}{\value{algosavedequation}}%
}

\allowdisplaybreaks[3]

\newcommand{\longthmtitle}[1]{\mbox{}\emph{(#1):}}

\newtheorem{assumption}{Assumption}
\newtheorem{problem}{Problem}

\newtheorem{remark}{Remark}
\newtheorem{theorem}{Theorem}
\newtheorem{definition}{Definition}
\newtheorem{corollary}{Corollary}
\newtheorem{proposition}{Proposition}
\newtheorem{lemma}{Lemma}

\newcommand{\real}{\mathbb{R}}

\newcommand{\naturalpos}{\mathbb{N}_{>0}}

\newcommand{\complex}{\mathbb{C}}

\newcommand*{\QEDBL}{\hfill\ensuremath{\blacksquare}}% Black square
\newcommand\oprocendsymbol{\hbox{$\square$}}
\newcommand\oprocend{\relax\ifmmode\else\unskip\hfill%
\fi\oprocendsymbol}

\DeclareMathAlphabet{\mymathbb}{U}{BOONDOX-ds}{m}{n}

\newcommand{\setdef}[2]{\{#1 \; : \; #2\}}

\newcommand{\bx}{{\mathbf{x}}}

\newcommand{\bu}{{\mathbf{u}}}
\newcommand{\bp}{{\mathbf{p}}}

\newcommand{\bv}{{\mathbf{v}}}

\newcommand{\Kc}{{\mathcal{K}}}
\newcommand{\Cc}{{\mathcal{C}}}

\newcommand{\norm}[1]{\Vert #1 \Vert}

\renewcommand{\baselinestretch}{.99}

\begin{document}

\title{\bf Characterization of the Dynamical Properties of Safety Filters \\ for Linear Planar Systems}

% Title: Stability Analysis of Safety Filters for Linear Planar Systems

% Title2: Characterization of Dynamical Properties of Safety Filters for Linear Planar Systems

\author{Yiting Chen$^*$ \quad Pol Mestres$^*$ \quad Emiliano Dall'Anese \quad Jorge Cort\'{e}s
\thanks{$^*$Equal contribution of the authors. Y. Chen and E. Dall'Anese are with the Department of Electrical and Computer Engineering at Boston University; P. Mestres and J. Cort\'{e}s are with the Department of Mechanical and Aerospace Engineering at the University of California San Diego.}
\thanks{This work was supported by the AFOSR Award FA9550-23-1-0740.}
}

\maketitle

\begin{abstract}%

This paper studies the dynamical properties of closed-loop systems obtained from control barrier function-based safety filters. We provide a sufficient and necessary condition for the existence of undesirable equilibria and show that the Jacobian matrix of the closed-loop system evaluated at an undesirable equilibrium always has a nonpositive eigenvalue. In the special case of linear planar systems and ellipsoidal obstacles, we give a complete characterization of the dynamical properties of the corresponding closed-loop system. We show that for underactuated systems, the safety filter always introduces a single undesirable equilibrium, which is a saddle-point. We prove that all trajectories outside the global stable manifold of such equilibrium converge to the origin.
In the fully actuated case, we discuss how the choice of nominal controller affects the stability properties of the closed-loop system. Various simulations illustrate our results.

% \blue{not sure if we need to mention the wierd behavior when the obstacle is not convex or not bounded in the Abstract.}
\end{abstract}

%\begin{keywords}%

%\end{keywords}

%%%%%%%%%%%%%%%%%%%%%%%%%%%%%%%%%%
\section{Introduction}
\label{sec:introduction}

Modern autonomous systems and cyber-physical systems -- from self-driving vehicles and robotic systems to critical infrastructures -- must provide safety guarantees while performing complex operational tasks~\cite{matni2024towards}. A popular approach to promote safety, where the term ``safety'' here refers to the ability to render a predefined set of states forward invariant, relies on the so-called \textit{safety filters}; these filters take a potentially unsafe nominal controller, designed to provide  stability or optimality guarantees, and minimally modify it to account for safety constraints. While the filtered controller ensures safety, it may not preserve the stability or optimality properties of the nominal controller. This challenge is the main motivation for this work.

\subsubsection*{Literature Review}
One of the main approaches for rendering a given  set forward invariant is via  Control Barrier Functions (CBFs)~\cite{PW-FA:07,ADA-SC-ME-GN-KS-PT:19,WX-CB:22,xiao2023safe}. Given a nominal controller with desirable properties such as  asymptotic stability of an equilibrium, CBFs acts on top of the nominal controller to ensure safety. This technique is often referred to as a \textit{safety filter}~\cite{LW-ADA-ME:17}. The main research question here is whether the closed-loop system with safety filters retains the stability guarantees of the nominal controller. This was studied in, e.g.,~\cite{WSC-DVD:22-tac}, which provides an estimate of the region of attraction of the equilibrium. However, it is unclear how conservative such estimate may be for general systems. The seminal works in~\cite{MFR-APA-PT:21,XT-DVD:21,PM-JC:23-csl,YY-SK-BG-NA:23,GN-SM:22} show that designs similar to safety filters can introduce undesirable equilibria that may be stable or unstable.

\subsubsection*{Statement of Contributions}
The goal of this paper is to advance the understanding of the dynamical properties of closed-loop systems obtained from CBF-based safety filters. %As previously noted in the literature, safety filters can fail to retain the global asymptotic stability properties of the nominal controller.
The main contribution of the paper is two-fold:

\noindent \emph{(i)} Our first contribution is to characterize the undesirable equilibria that emerge in the closed-loop system formed by a control-affine dynamical system, a stabilizing nominal controller, and a CBF-based safety filter. General obstacles are considered (this is the subject of Section~\ref{sec:general-results}).

\noindent \emph{(ii)} Next, we focus our attention to linear time-invariant (LTI) planar systems (Section~\ref{sec:lti-planar-circular-obstacles}). We show that, for these systems, the dynamical properties of systems with ellipsoidal obstacles are equivalent to those of systems with circular obstacles. For underactuated LTI planar systems, we give a complete characterization of the trajectories of the closed-loop system. We show that such systems always have a single undesirable equilibrium. Moreover, we show that such undesirable equilibrium is a saddle point and show that all trajectories that lie outside the global stable manifold of this equilibrium converge to the origin. For fully actuated LTI planar systems, we show that the closed-loop system can have up to three undesirable equilibria, and characterize their stability properties. 

Additionally, we show that in the fully actuated case there always exists a nominal controller (which can be explicitly computed) that makes the closed-loop system have a single undesirable saddle point equilibrium. Therefore, our findings can be used to inform the design of the nominal controller.
For reasons of space, proofs are included in an extended version~\cite{extended}.

%%%%%%%%%%%%%%%%%%%%%%%%%%%%%%%%%%
\section{Preliminaries and Problem Statement}
\label{sec:prelim}

\emph{Notation}. We denote by $\naturalpos$ and $\real$ the set of positive integers, real, and nonnegative numbers.  
We use bold symbols to represent vectors and non-bold symbols to represent scalar quantities; $\textbf{0}_n$ represents the $n$-dimensional zero vector.
Given $\bx\in\real^{n}$, $\norm{\bx}$ denotes its Euclidean norm. Given a matrix $G\in\real^{n\times n}$, $\norm{\bx}_{G} = \sqrt{\bx^T G \bx}$. A function 
$\beta:\real\to\real$ is of extended class $\mathcal{K}_{\infty}$ if $\beta(0)=0$,
$\beta$ is strictly increasing and $\lim\limits_{s\to\pm\infty}\beta(s)=\pm\infty$.
Given a set $S\subset\real^n$, we denote by $\text{Int}(S)$ and $\partial S$ the interior and boundary of $S$, respectively. For a continuously differentiable function  $h:\real^{n}\to\real$, $\nabla h(\bx)$ denotes its gradient at $\bx$.  %A function $f:\real^n\to\real^n$ is coercive if $\lim\limits_{\norm{x}\to\infty} \frac{f(x)^T x}{\norm{x}}=\infty$.

Consider the system $\dot{\bx} = f(\bx)$, with $f:\real^n\to\real^n$ locally Lipschitz. %Then, given $\bx_0\in\real^n$ we let $\bx(t;\bx_0)$ be the locally unique solution of the dynamical system with initial condition at $\bx_0$.
Then, for any initial condition $\bx_0 \in \real^n$ at time $t_0$, there exists a maximal interval of existence $[t_0,t_1)$ such that $\bx(t;\bx_0)$ is the unique solution to $\dot{\bx} = f(\bx)$ on 
$[t_0,t_1)$, cf.~\cite{sontag2013mathematical}.
For $f$  continuously differentiable
and $\bx^*$ an equilibrium point of $f$ (i.e., $f(\bx^*)=\textbf{0}_n$), $\bx^*$ is \emph{degenerate} if the Jacobian of $f$ evaluated at $\bx^*$ has at least one eigenvalue with real part equal to zero (otherwise, we refer to  $\bx^*$ as \emph{hyperbolic}).
%
%\marginJC{eigenvalue equal to zero or with real part equal to zero? \blue{YC: yes, it should be  with real part equal to zero.}}
%
% We say that $\bx^*$ is hyperbolic if the Jacobian of $f$ evaluated at $\bx^*$ has no eigenvalues with zero real part.
Given a hyperbolic equilibrium point with $k\in\mathbb{Z}_{>0}$ eigenvalues with negative real part, the Stable Manifold Theorem~\cite[Section 2.7]{LP:00} ensures that there exists an invariant $k$-dimensional manifold $S$ for which all trajectories with initial conditions lying on $S$ converge to $\bx^*$.
The global stable manifold at $\bx^*$ is defined as $W_s(\bx^*)=\bigcup\limits_{ \{t\leq0, \ \bx_0\in S \} }\bx(t;\bx_0)$. Given a complex number $z\in\complex$, $\text{Re}(z)$ denotes its real part.

\subsection{Control barrier functions and safety filters} 

Consider a control-affine dynamical system of the form
\begin{align}\label{eq:control-affine-sys}
  \dot{\bx}=f(\bx)+g(\bx)\bu,
\end{align}
where $f:\real^{n} \! \to \! \real^{n}$ and $g:\real^{n} \! \to \!  \real^{n\times m}$
are locally Lipschitz functions, $\bx\in\real^{n}$ is the state, and
$\bu\in\real^{m}$ is the input. 
% The following is a  definition of CBF for the system~\eqref{eq:control-affine-sys}. 

\vspace{.1cm} 

\begin{definition}[Control Barrier
    Function]\label{def:cbf}
  Let $h:\real^{n}\to\real$ be a continuously differentiable function, and define 
  the set $ \mathcal{C}=\setdef{\bx\in\real^n}{h(\bx)\geq0}$.  The
  function $h$ is a \textbf{CBF} of $\Cc$ for the
  system~\eqref{eq:control-affine-sys} if there exists an extended class
  $\mathcal{K}_{\infty}$ function $\alpha$ such that, for all $\bx \in \Cc$, there exists $\bu \in\real^{m}$ satisfying $\nabla h(\bx)^\top (f(\bx)+g(\bx)\bu) + \alpha(h(\bx)) \geq 0$. \hfill $\Box$
\end{definition}

\vspace{.1cm} 

Suppose that a nominal controller $\bu=k(\bx)$ is designed so that the system 
$\dot \bx = \tilde{f}(\bx) := f(\bx)+g(\bx)k(\bx)$ renders the  origin globally asymptotically stable (this is  without loss of generality). Consider the system
\begin{equation}\label{eq:general-system-1}
\dot{\bx}= \tilde{f}(\bx) +g(\bx)v(\bx),
\end{equation}
 where the map $\bx \mapsto v(\bx)$ is defined as:
\begin{align} \label{eq:v-problem} 
%\begin{array}{rl}
v(\bx)&=\arg\underset{ \boldsymbol{\theta} \in \mathbb{R}^{ m} }{\min}  \| \boldsymbol{\theta} \|_{G(\bx)}^2 \\
\notag
&\text { s.t. } \nabla h(\bx)^\top (f(\bx)+g(\bx)(k(\bx)+ \boldsymbol{\theta} ))+\alpha( h(\bx)) \geq 0  
%\end{array}
\end{align}
with $G:\real^n\to\real^{m\times m}$  continuously differentiable and positive definite for all $\bx\in\real^n$. We assume the following. 

%\red{YC: we should assume the origin is the eq of $\dot x=f(x)+g(x)k(x)$, right?}

\vspace{.1cm}

\begin{assumption}[Origin in the interior of $\mathcal{C}$]\label{as: interior eq}
    The set $\{ \bx\in\real^n :~h(\bx)= 0,~\tilde{f}(\bx)=\textbf{0}_n\}$ is empty and $h(\textbf{0}_n)>0$. \hfill $\Box$
\end{assumption}

\vspace{.1cm}

\begin{assumption}[Feasibility]\label{as: feasibility}
    There exists an extended class
  $\mathcal{K}_{\infty}$ function $\alpha$ such that $g(\bx)^\top \nabla h(\bx)\neq 0$ for all $\bx$ in $\{\bx\in\real^n : h(x)\geq 0,~\nabla h(\bx)^\top f(\bx)+\alpha( h(\bx)) \leq 0  \}$. \hfill $\Box$
\end{assumption}

\vspace{.1cm}

Assumption~\ref{as: feasibility} ensures that~\eqref{eq:v-problem} is feasible for all $\bx\in\real^n$ and therefore $v(\bx)$ is well-defined for all $\bx\in\real^n$.
Moreover, under Assumption~\ref{as: feasibility}, and using  arguments similar to~\cite[Lemma III.2]{MA-NA-JC:23-tac}, one can show that $v(\bx)$ is locally Lipschitz. 
Assumption~\ref{as: feasibility} also ensures that $\frac{\partial h}{\partial \bx}(\bx) \neq \textbf{0}_n$ for all $\bx\in\partial\Cc$.
From~\cite[Thm.~2]{ADA-SC-ME-GN-KS-PT:19}, it follows that the system~\eqref{eq:general-system-1} with the controller $v(\bx)$ renders the set $\Cc$ forward invariant. Because of this feature, and because $\Cc$ is modeling a set of \emph{safe} states, $v(\bx)$ is typically referred to as \emph{safety filter}.

%%%%%%%%%%%%%%%%%%%%%%%%%%%%%%%%%%
\subsection{Problem Statement}
\label{sec:problem-statement}

We consider a control-affine dynamical system as in~\eqref{eq:control-affine-sys} and a safe set $\Cc\subset\real^n$ defined as the $0$-superlevel set of a differentiable function $h:\real^n\to\real$. Assume that $h$ is a CBF of $\Cc$ for system~\eqref{eq:control-affine-sys}, and that Assumptions~\ref{as: interior eq} and~\ref{as: feasibility} hold. Studying the dynamical behavior of~\eqref{eq:general-system-1} is challenging.
Indeed, as noted in~\cite{WSC-DVD:22-tac}, it does not inherit the global asymptotic stability properties of the controller $k$, and can even have undesirable equilibria~\cite{MFR-APA-PT:21,XT-DVD:21,PM-JC:23-csl,YY-SK-BG-NA:23}. However, most of these works focus on studying conditions under which such undesirable equilibria exist or can be confined to specific regions of interest, but do not study dynamical properties of the closed-loop system. Hence the goal of this paper is as follows:

\vspace{.1cm}

% \begin{problem}
% Given the system~\eqref{eq:control-affine-sys} with a stabilizing nominal controller $k(\bx)$ and the safety filter $v(\bx)$,  characterize the equilibria of the closed-loop system along with their stability properties and regions of attraction.   \hfill $\Box$ 

% \end{problem}

\begin{problem}\label{problem}
Given system~\eqref{eq:control-affine-sys} with a stabilizing nominal controller $k(\bx)$ and the safety filter $v(\bx)$,  characterize the dynamical properties of \eqref{eq:general-system-1} (such as undesirable equilibria and their regions of attraction, limit cycles and region of attraction of the origin) and investigate how these properties are determined by the original closed-loop system $\dot \bx=f(\bx)+g(\bx)k(\bx)$.   \hfill $\Box$ 

\end{problem}

\vspace{.1cm}

In the following section, we consider the  system~\eqref{eq:general-system-1} and characterize its undesirable equilibria. In Section~\ref{sec:lti-planar-circular-obstacles}, given the complexity of solving Problem~\ref{problem}, we then restrict our attention to linear planar systems.
%
% \marginJC{Why don't we say here that the majority of the state of the art seeks to identify conditions that rules out undesired equilibria or confines them to specific areas, whereas here our viewpoint is  more general and we seek to characterize their dynamical properties. Given the complexity of the problem, this is why we restrict our attention to planar systems.}
%
%%%%%%%%%%%%%%%%%%%%%%%%%%%%%%%%%%
\section{Characterization of undesirable equilibria}\label{sec:general-results}

We start by reformulating the expression for the unique optimal solution $v(\bx)$ of the quadratic program~\eqref{eq:v-problem}. Let $\eta(\bx) = \nabla h(\bx)^T (f(\bx)+g(\bx)k(\bx)) + \alpha(h(\bx))$. Then,
\begin{align}\label{eq:v-expression}
    v(\bx) = \begin{cases}
        \textbf{0}_m, &\ \text{if} \ \eta(\bx) \geq 0, \\
        \bar{u}(\bx), &\ \text{if} \ \eta(\bx) < 0,
    \end{cases}
\end{align}
where $\bar{u}(\bx):=-\frac{\eta(\bx) G(\bx)^{-1}g(\bx)^\top\nabla h(\bx) }{\| 
 g(\bx)^\top \nabla h(\bx)  \|_{ { G(\bx)^{-1}} }^2}$. We use this expression in the following result, which provides a necessary and sufficient condition for undesirable equilibria of~\eqref{eq:general-system-1}. 

\vspace{.1cm}

%
% \marginJC{Is this result ours? I'd think this is in the prior literature. \blue{YC: I can find a similar result for CLF-CBF-QP framework, but I didn't find a result for safe filter. I agree that they are close related.}}
%
% \marginJC{Maybe then point out that a similar result can be found in bla, bla}
%
\begin{lemma}\longthmtitle{Conditions for undesirable equilibria}\label{lem:undesired-eq-characterization}
  Let Assumptions \ref{as: interior eq}  and \ref{as: feasibility} be satisfied. Let $\bp_0 \in \mathbb{R}^n$ be such that $\tilde{f}(\bp_0) \neq \textbf{0}_n$. Then,  $\bp_0$ is an equilibrium of \eqref{eq:general-system-1}
    if and only if  there exists $\delta<0$ such that
    \vspace{-.1cm}
\begin{align}
    ~~~~~~~ & h(\bp_0)=0 ~\text{and} \label{eq: condition-eq} & \\ 
    ~~~~~~~ & \tilde{f}(\bp_0)=\delta g(\bp_0)G(\bp_0)^{-1}g(\bp_0)^\top\nabla h(\bp_0) \, . \nonumber & ~~~~~~~~~ \Box
\end{align}
\end{lemma}
\vspace{.1cm}

\vspace{.1cm}

% \begin{lemma}[]
%     Let Assumption \ref{as: interior eq} and \ref{as: feasibility} be satisfied.  Given that $\bp_0$ satisfying that $\tilde{f}(\bp_0) \neq \textbf{0}_n$, then  $\bx=\bp_0$ is an equilibrium of \eqref{eq:general-system-1}
%     if and only if  
%     there exists $\delta<0$ such that $(\bp_0,\delta)$ is a solution of \eqref{eq: condition-eq} .
% \end{lemma}

This result has the same flavor as~\cite[Theorem 2]{XT-DVD:21} and~\cite[Proposition 5.1]{PM-JC:23-csl}, which characterize the undesirable equilibria for related, but different, safety filter designs.

By Lemma \ref{lem:undesired-eq-characterization}, we can define the set of \emph{potential undesirable equilibria} of \eqref{eq:general-system-1} as: 
    $$\mathcal{E}:=\{\bx:~\exists~\delta\in\mathbb{R} \ \text{s.t. $(\bx,\delta)$ solves \eqref{eq: condition-eq}}      \} \, .$$
On the other hand, the set of \emph{undesirable equilibria} is: 
\[ \hat{\mathcal{E}}:=\{\bx:~\exists~\delta<0  \ \text{s.t. $(\bx,\delta)$ solves \eqref{eq: condition-eq}}\} \subset \mathcal{E}.\]

{ The term \textit{undesirable} stems from the fact that these equilibria are different from the origin, which is the equilibrium point where the system needs to be stabilized.}
By Lemma \ref{lem:undesired-eq-characterization}, it follows that determining the equilibrium points of system \eqref{eq:general-system-1} is equivalent to solving \eqref{eq: condition-eq} and checking the sign of $\delta$. For a solution $(\bp_0,\delta_{\bp_0})$ to \eqref{eq: condition-eq}, we refer to  $\delta_{\bp_0}$ as the \textbf{indicator} of $\bp_0$, since the sign of $\delta_{\bp_0}$
determines whether $\bp_0$ is a new,  \textit{undesirable}, equilibrium of the  system with the CBF filter. Additionally, we  show that the value of the indicator is useful for determining the stability properties of the undesirable equilibrium. For a given undesirable equilibrium $\bp_0$ of \eqref{eq:general-system-1}, the indicator can be computed as $\delta_{\bp_0}=\frac{\nabla h(\bp_0)^\top \tilde{f}(\bp_0)    }{\|  g(\bp_0)^\top \nabla h(\bp_0)  \|_{ { G(\bx)^{-1}} }^2} $. In addition, Assumption \ref{as: interior eq} ensures that no solution of \eqref{eq: condition-eq} has $\delta=0$.

Under appropriate conditions, the next result shows that we can compute the Jacobian of $\tilde{f}(\bx)+g(\bx)v(\bx)$ at $\bx\in\hat{\mathcal{E}}$ and find one of its eigenvalues.

\vspace{.1cm}

\begin{lemma}\longthmtitle{Jacobian at the undesirable equilibrium}\label{lem:jacobian-characterization-general} Let Assumptions \ref{as: interior eq} and \ref{as: feasibility} be satisfied and assume that $D = g(\bx){ G(\bx)^{-1}}g(\bx)^\top$ is a constant matrix, $\tilde{f}(\bx)$, $\alpha(\cdot)$ are differentiable and $h(\bx)$ is twice differentiable. For any $\bx\in\hat{\mathcal{E}}$, the Jacobian of $\tilde{f}(\bx)+g(\bx)v(\bx)$ evaluated at $\bx$ is 
\begin{align*}
&J\mid_{\bx\in\hat{\mathcal{E}}}=J_{\Tilde{f}}-\frac{ D \nabla h(\bx)\nabla h(\bx)^\top}{\nabla h(\bx)^\top D \nabla h(\bx)  } [ J_{\Tilde{f}}+\alpha^\prime(0)\mathbf{I}_n ]\\
    &- \frac{D}{\nabla h(\bx)^\top D \nabla h(\bx) }  [ H_{h} \nabla h(\bx)^\top \Tilde{f}(\bx)-  \nabla h(\bx) \Tilde{f}(\bx)^\top H_h],
\end{align*}
where $J_{\Tilde{f}}$ is the Jacobian matrix of $\tilde{f}(\bx)$ and $H_h$ is the Hessian of $h(\bx)$. 
Moreover, for any $\bx\in\hat{\mathcal{E}}$, it holds that
$$ (J \mid_{\bx\in\hat{\mathcal{E}}})^\top \nabla h(\bx)=-\alpha^\prime(0) \nabla h(\bx),$$
the algebraic multiplicity of $-\alpha^\prime(0)$ is $1$, and all the other eigenvalues of $J\mid_{\bx\in\hat{\mathcal{E}}}$ %is independent of  $\alpha(\cdot)$. 
do not change when $\alpha(\cdot)$ changes. \hfill $\Box$
\end{lemma}

\vspace{.1cm}
The proof of Lemma~\ref{lem:jacobian-characterization-general} follows from a careful computation. Note that $J$ always has an eigenvalue $-\alpha^\prime(0)$; it follows that all the undesirable equilibria are degenerate  
if $\alpha^\prime(0) = 0$, which complicates the stability analysis. If $\alpha^\prime(0) >0$, the Jacobian evaluated at  $\bx\in\hat{\mathcal{E}}$ always has a negative eigenvalue.
%(we recall that an equilibrium is degenerate if  the Jacobian evaluated at that point is not full rank, see \cite{JM-book:07}). 
% Therefore, we impose the following assumption.
% \vspace{.1cm}
% \begin{assumption}\label{as: d_alpha(0)>0}
%     The extended $\mathcal{K}_\infty$ function  $a \mapsto \alpha(z)$ is differentiable 
%     and $\alpha^\prime(0)>0$. \hfill $\Box$
% \end{assumption}
% \vspace{.1cm}
Lemmas \ref{lem:undesired-eq-characterization} and \ref{lem:jacobian-characterization-general} show that the extended $\mathcal{K}_\infty$ function
$\alpha(\cdot)$ does not play a role in the existence of undesirable equilibria. Additionally, changing $\alpha(\cdot)$ will only affect one eigenvalue of the Jacobian evaluated at $\bx\in\hat{\mathcal{E}}$. 
The assumption that $g(\bx){ G(\bx)^{-1}}g(\bx)^\top$ is constant is satisfied for several classes of systems, such as mechanical systems, like the ones considered in~\cite[Section III.B]{WSC-DVD:22-tac}.

%%%%%%%%%%%%%%%%%%%%%%%%%%%%%%%%%%
\section{LTI Planar Systems with Safety Filters}
\label{sec:lti-planar-circular-obstacles}

Since Problem~\ref{problem} is difficult to solve in general, here we provide a solution for it for planar LTI dynamics and ellipsoidal obstacles.
Consider the LTI planar system
\begin{align}\label{eq:2d-linear-system}
        \dot{\bx} = A\bx + B\bu,
\end{align}
with $\bx = [x_1,x_2]^\top \in\real^2$, $\bu\in\real^m$, with $m\in\{1,2\}$, $A\in\real^{2\times 2}$, and with $B\in\real^{2 \times m}$ full column rank. We make the following assumption on~\eqref{eq:2d-linear-system}.

\vspace{.1cm}

\begin{assumption}[Stabilizability]\label{as:A-B-stabilizable}
    The system~\eqref{eq:2d-linear-system} is stabilizable. Moreover, let $\bu = - K \bx$, $K \in\real^{2\times m}$, be any stabilizing controller such that $\tilde{A} = A-BK$ is Hurwitz.  \hfill $\Box$
\end{assumption}

\vspace{.1cm}

In this setup, the system \eqref{eq:general-system-1} is then customized as follows: 
    \begin{align}\label{eq:v-linear-system}
        \dot{\bx} = F(\bx) := (A-BK)\bx + B v(\bx),
    \end{align}
    where the safety filter is given by
    \begin{align}\label{eq:v-linear-expression}
    v(\bx) = \begin{cases}
                0, &\ \text{if} \ \eta(\bx) \geq 0, \\
                -\frac{\eta(\bx) G(\bx)^{-1} B^T\nabla h(\bx) }{ 
            \norm{B^T \nabla h(\bx)}_{ { G(\bx)^{-1}} }^2 }, &\ \text{if} \ \eta(\bx) <  0 . 
            \end{cases}
    \end{align}

In the following, we show that the undesirable equilibria and their stability properties of \eqref{eq:v-linear-system} with ellipsoidal obstacles are equivalent to those of a system with circular obstacles.

\vspace{.1cm}

\begin{proposition}\longthmtitle{Safety filters with ellipsoidal and circular obstacles have the same dynamical properties}\label{prop:ellipsoidal-circular-same-properties}
    Let $\bx_c\in\real^2$, $P\in\real^{2 \times 2}$ positive definite, $h(\bx) = (\bx-\bx_c)^T P (\bx-\bx_c) - 1$, $\Cc:=\setdef{\bx\in\real^n}{h(\bx)\geq0}$.
    Suppose that $P=E^T E$, with $E\in\real^{2\times 2}$ also positive definite, and define $\hat{\bx}_c = E \bx_c$, 
    $\hat{h}(\hat{\bx}) = (\hat{\bx}-\hat{\bx}_c)^T (\hat{\bx}-\hat{\bx}_c) - 1$ and $\hat{\Cc} = \setdef{\bx\in\real^n}{\hat{h}(\bx)\geq0}$. Moreover, let $\hat{A} = EAE^{-1}$, $\hat{B}=E B$, $\hat{G}(\hat{\bx}) = G(E^{-1}\hat{\bx})$ and $\hat{\eta}(\hat{\bx}) = \nabla \hat{h}(\hat{\bx})^T (\hat{A}-\hat{B}K E^{-1})\hat{\bx}+\alpha(\hat{h}(\hat{\bx}))$.
    Consider the system
    \begin{align}\label{eq:vhat-linear-system}
        \dot{\hat{\bx}} = \hat{F}(\hat{\bx})  :=(\hat{A}-\hat{B}KE^{-1})\hat{\bx} +  \hat{B} \hat{v}(\hat{\bx}),
    \end{align}
    where 
    \begin{align}\label{eq:vhat-linear-expression}
    \hat{v}(\hat{\bx}) = \begin{cases}
                0, &\ \text{if} \ \hat{\eta}(\hat{\bx}) \geq 0, \\
                -\frac{\hat{\eta}(\hat{\bx}) \hat{G}(\hat{\bx})^{-1}(\hat{\bx}) \hat{B}^T \nabla \hat{h}(\hat{\bx}) }{ \norm{
            \hat{B}^T \nabla \hat{h}(\hat{\bx})}_{\hat{G}(\hat{\bx})^{-1}}^2 }, &\ \text{if} \ \hat{\eta}(\hat{\bx}) < 0
            \end{cases}
    \end{align}
   
    Then, 
    \begin{enumerate}[i)]
        \item\label{it:ellip-first} $\hat{\Cc}$ is forward invariant under system~\eqref{eq:vhat-linear-system} and $\Cc$ is forward invariant under system~\eqref{eq:v-linear-system};
        \item\label{it:ellip-lipschitz} systems~\eqref{eq:vhat-linear-system} and~\eqref{eq:v-linear-system} are locally Lipschitz;
        \item\label{it:ellip-second} $(A,B)\!$ is stabilizable if and only if $(\hat{A},\hat{B})$ is stabilizable;
        \item\label{it:ellip-third} $\hat{\bp}\in\real^2$ is an undesirable equilibrium of~\eqref{eq:vhat-linear-system} if and only if $\bp := E^{-1} \hat{\bp}$ is an undesirable equilibrium of~\eqref{eq:v-linear-system};
        \item\label{it:ellip-fourth} 
        % the Jacobian of $\hat{F}$ at $\hat{\bp}$ has the same number of positive, negative and zero eigenvalues as the Jacobian of $F$ at $\bp$. 
        the Jacobian of $\hat{F}$ at $\hat{\bp}$ and the Jacobian of $F$ at $\bp$ are similar. 
        \hfill $\Box$
    \end{enumerate}
\end{proposition}

\vspace{.1cm}

Given that Proposition~\ref{prop:ellipsoidal-circular-same-properties} ensures that undesirable equilibria for general ellipsoidal obstacles have the same stability properties as undesirable equilibria for circular obstacles, %and given that calculations are simpler with circular obstacles, 
in the following we focus on studying the dynamical properties of safety filters for LTI systems and circular obstacles.

Accordingly, we consider the circular unsafe set:
\begin{align*}
    \Cc = \setdef{ \bx \in\real^2}{ h(\bx) = \norm{\bx-\bx_c}^2 - r^2 \geq 0},
\end{align*}
with $\bx_c \in\real^2$ the center. We take the extended class $\Kc_\infty$ function in Definition~\ref{def:cbf} to be linear and with slope $\alpha_0>0$.  We denote the eigenvalues of $\Tilde{A}$ as $\lambda_1$, $\lambda_2\in\mathbb{C}^2$. Let $V(\bx)=\bx^\top Q \bx$ be the associated Lyapunov function, with a positive definite symmetric matrix $Q$, such that   $\bx^\top Q \tilde{A} \bx<0$  for all $\bx\neq \textbf{0}_2$.  Additionally, we pick $G(\bx)=B^\top B$. 

The first result rules out the existence of limit cycles.

\vspace{.1cm}

\begin{proposition}\longthmtitle{Non-existence of limit cycles}\label{prop:no-limit-cycles}
    %Suppose that the assumptions in Lemma~\ref{lem:interior} and Proposition~\ref{prop:cbf} hold. 
    % comment by YC: Lemma~\ref{lem:interior} is equivalent to assumption  \ref{as: interior eq}
    % need to double check Proposition~\ref{prop:cbf}    
   Suppose that Assumptions \ref{as: interior eq}--\ref{as:A-B-stabilizable} hold for the closed-loop system~\eqref{eq:v-linear-system}. { Assume that for~\eqref{eq:v-linear-system}}, $\hat{\mathcal{E}}=\{\hat{\bx}^*\}$, with $ \hat{\bx}^*$ a saddle point.
   %
%\marginJC{DO we have any results for when $\hat{\mathcal{E}}$ is not a singleton? Or will it always be a singleton in the planar case? If so, this point is not clearly stated}
%\marginPM{$\hat{\mathcal{E}}$ is not always a singleton in the planar case. We could give a similar result if the sum of indices of the points in $\hat{\mathcal{E}}$ is $\leq -1$ but the result would look a bit more obscure.}
%\marginYC{We can also add what Pol says as a comment, like "if the sum of indices of the points in $\hat{\mathcal{E}}$ is $\leq -1$ and there is no degenerate equilibrium......". So we can also conclude that when there is one undesired asymptotically stable eq and two saddle points, there is no limit cycle. However, we still cannot have the result for the general case, since When there is a degenerate equilibrium, index theorem is gone.   } %% Let's leave it for the journal version!
   %
    Then, there exist $\alpha_1^*>0$ such that for any $\alpha(s)=\alpha_0 s$ with $\alpha_0\geq \alpha_1^*$,~\eqref{eq:v-linear-system} does not have limit cycles in $\mathcal{C}$. \hfill $\Box$
    % for any extended class $\Kc$ function $\alpha$ and stabilizing controller $k$, the closed-loop system~\eqref{eq:general-system-1} obtained from~\eqref{eq:2d-linear-system} does not have limit cycles not circling the origin or circling the origin and $\real^n\backslash\Cc$.
    % Moreover, let $\Rc\subset\real^2$ be a compact set with $\textbf{0}_2\in\Rc$.
    % Let $\bv$ be an eigenvector of the Hurwitz matrix $A-BK$ with eigenvalue $\lambda$. Further let $c\in\real$ be such that
    % $c \bv \notin\Rc$ and $S:=\setdef{\theta c \bv}{\theta\in[0,1]}$
    % is such that $S\subset \text{Int}(\Cc)$.
    % Then, by taking $\alpha:\real\to\real$ with $\alpha(s)=\alpha_0 s$, and
    % \begin{align}\label{eq:alpha-condition-no-limit-cycles}
    %     \alpha_0 > \sup_{t\geq0} \frac{|(c \bv e^{\lambda t}-\bx_c)^T(A-BK)c \bv e^{\lambda t} |}{h(c \bv e^{\lambda t})},
    % \end{align}
    % the closed loop system~\eqref{eq:general-system-1} obtained from system~\eqref{eq:linear-2d-underactuated-system} does not have any limit cycles contained in $\Rc$ and circling the origin.
\end{proposition}

\vspace{.1cm}

By combining the results in this section, we have the following.

\vspace{.1cm}

\begin{theorem}[Global behavior analysis]\label{thm:complete-characterization-planar-linear-underactuated}
Suppose that the Assumptions \ref{as: interior eq}--\ref{as:A-B-stabilizable} hold for the closed-loop system~\eqref{eq:v-linear-system}. Assume that $\hat{\mathcal{E}}=\{\hat{\bx}^*\}$ and $ \hat{\bx}^*$ is a saddle point.
Then, there exists $\alpha_2^*>0$ such that for any $\alpha(s)=\alpha_0 s$ with $\alpha_0\geq \alpha_2^*$, if $W_s(\hat{\bx}^*)$ denotes the global stable manifold of $\hat{\bx}^*$ it holds that:
    \begin{enumerate}
        \item if $\bx_0 \in W_s(\hat{\bx}^*)$, then $\lim\limits_{t\to\infty} \bx(t;\bx_0) = \hat{\bx}^*$;
        \item if $\bx_0 \notin W_s(\hat{\bx}^*)$, then $\lim\limits_{t\to\infty} \bx(t;\bx_0) = \textbf{0}_2$. \hfill $\Box$
    \end{enumerate}

    % Suppose that assumptions in Lemma~\ref{lem:interior} and Proposition~\ref{prop:cbf} hold.
    % Let $\Rc\subset\real^2$ be a compact set.
    % Let $\bx_0\in\Rc$ and let $\bx(t;\bx_0)$ denote the solution of~\eqref{eq:general-system-1} for system~\eqref{eq:2d-linear-system} at time $t$ with initial condition at $\bx_0$.
    % Then, there exists $\alpha_{\Rc}>0$ such that for all class $\Kc$ functions $\alpha(s)=\alpha_0 s$ with $\alpha_0 >\alpha_{\Rc}$,
    % \begin{enumerate}
    %     \item\label{it:first} if $\gamma x_c + \beta y_c < 0$ and $\bx_0$ belongs to the stable manifold of $\bp_{+}$, then $\lim\limits_{t\to\infty} \bx(t;\bx_0) = \bp_{+}$;
    %     \item\label{it:second} if $\gamma x_c + \beta y_c < 0$ and $\bx_c$ does not belong to the stable manifold of $\bp_{+}$, then $\lim\limits_{t\to\infty} z(t;\bx_0) = \textbf{0}_2$;
    %     \item\label{it:third} if $\gamma x_c + \beta y_c > 0$ and $\bx_0$ belongs to the stable manifold of $\bp_{-}$, then $\lim\limits_{t\to\infty} \bx(t;\bx_0) = \bp_{-}$;
    %     \item\label{it:fourth} if $\gamma x_c + \beta y_c > 0$ and $x_0$ does not belong to the stable manifold of $\bp_{-}$, then $\lim\limits_{t\to\infty} \bx(t;\bx_0) = \textbf{0}_2$.
    % \end{enumerate}
\end{theorem}

\vspace{.1cm}

%\begin{remark}\longthmtitle{Characterization of trajectories for linear systems with ellipsoidal obstacles}%\label{rem:characterization-trajectories-ellipsoidal}
%    Proposition~\ref{prop:ellipsoidal-circular-same-properties} shows that there is a bijection between the undesired equilibria of a system of the form~\eqref{eq:v-linear-system} with ellipsoidal obstacles and the undesired equilibria of a system of the form~\eqref{eq:v-linear-system} with circular obstacles. Moreover, it shows that the stability properties of such undesired equilibria are the same. Hence, an analogue of Theorem~\ref{thm:complete-characterization-planar-linear-underactuated} for ellipsoidal obstacles holds. \hfill $\Box$
%\end{remark}

\vspace{.1cm}

\begin{remark}[Almost global asymptotic stability]
The Stable Manifold Theorem~\cite[Ch. 2.7]{LP:00} ensures that if $\hat{\bx}^*$ is a saddle point in $\real^2$, the local stable manifold is $1$-dimensional.
Therefore, it has measure of zero. 
Moreover, the global stable manifold
%Which might not be a manifold!
must also have measure of zero.
If this were not the case, solutions would have to intersect. However this is not possible 
due to the uniqueness of solutions. Hence $\setdef{\bx_0\in\real^n}{\lim\limits_{t\to\infty}\bx(t;\bx_0) = \textbf{0}_n} = S$.
It follows that the set of initial conditions whose associated trajectory converges to $\hat{\bx}^*$ has measure zero. \hfill $\Box$
\end{remark}

%\vspace{.1cm}

%Next, we focus on the existence and stability properties of the undesirable equilibria.

\subsection{Under-actuated LTI Planar Systems}

% \red{[ED: careful, we do not write anywhere that we denote as $x$ and $y$ the coordinates. I would advocate for $x_1$ and $x_2$]}

In the under-actuated case, we write
\begin{align}\label{eq:linear-2d-underactuated-system}
  A=\begin{bmatrix}
        a_{11} & a_{12} \\
        a_{21} & a_{22}
    \end{bmatrix},~
    B=
    \begin{bmatrix}
        b_1 \\
        b_2
\end{bmatrix},~\bx=\begin{bmatrix}
        x_{1}\\x_{2}
    \end{bmatrix},
\end{align}

% system~\eqref{eq:2d-linear-system} can be re-written as
% \begin{align}\label{eq:linear-2d-underactuated-system}
%     \begin{pmatrix}
%         \dot{x} \\ \dot{y}
%     \end{pmatrix} = \underbrace{\begin{pmatrix}
%         a_{11} & a_{12} \\
%         a_{21} & a_{22}
%     \end{pmatrix}}_{A}
%     \begin{pmatrix}
%         x \\
%         y
%     \end{pmatrix} + 
%     \underbrace{
%     \begin{pmatrix}
%         b_1 \\
%         b_2
%     \end{pmatrix}}_{B} u.
% \end{align}

Throughout this section, we denote $\bx_c = [x_{c,1}, x_{c,2}]^\top$ 
%the center of the circular obstacle
and let $\beta=a_{11}b_2 - b_1 a_{21}$, $\gamma = a_{22} b_1 - b_2 a_{12}$, and $T_3 = -\gamma x_{c,2} + \beta x_{c,1}$ and assume that $k:\real^2\to\real$ is a linear stabilizing controller of the form $k(\bx)=-K\bx=-k_1 x_1 - k_1 x_2$ for some $k_1, k_2 \in \real$. We note also that since in this case $G$ is a scalar,~\eqref{eq:v-linear-system} is independent of $G$.

The following results give conditions on $h$ and system~\eqref{eq:linear-2d-underactuated-system} that ensure that Assumptions~\ref{as: interior eq} and~\ref{as: feasibility} hold. 

\vspace{.1cm}

\begin{lemma}[Conditions for Assumption~\ref{as: interior eq}]\label{lem:interior}
   Assumption~\ref{as: interior eq} holds if and only if
     $\norm{\bx_c}^2 > r^2$.
     \hfill $\Box$
\end{lemma}

\vspace{.1cm}

The proof of Lemma~\ref{lem:interior} follows from the observation that $\norm{\bx_c}^2 > r^2$ guarantees that the origin is safe.

\vspace{.1cm}

\begin{proposition}[Conditions for Assumption~\ref{as: feasibility}]
\label{prop:cbf}
    Let $\alpha_0>0$, $T_1 := b_2 \beta + b_1 \gamma + \frac{1}{2} \alpha_0 (b_2^2 + b_1^2)$, and $T_2 := (\beta x_{c,1} - \gamma x_{c,2})^2 + 2\alpha_0 r^2 T_1$.
    Suppose that $r>0$, $b_1^2 + b_2^2 > 0$, $T_1>0$, and $\frac{r}{ \sqrt{b_2^2 + b_1^2} } > \frac{  |T_3| + \sqrt{ T_2 } }{ 2 T_1 }$.
    % \begin{align*}
    %     \frac{r}{ \sqrt{b_2^2 + b_1^2} } > \frac{  |T_3| + \sqrt{ T_2 } }{ 2 T_1 }.
    % \end{align*}
    Then, Assumption~\ref{as: feasibility} holds with the linear extended class $\Kc$ function $\alpha(s)=\alpha_0 s$. \hfill $\Box$
\end{proposition}

% Throughout the rest of this section, we take the extended class $\Kc$ function in Definition~\ref{def:cbf} to be linear and with slope $\alpha_0>0$.

\vspace{.1cm}

We next give a result that will be used later in the paper.

\vspace{.1cm}

\begin{lemma}[Conditions for 
$\beta$ and $\gamma$]\label{lem:discriminant-is-positive-if-cbf}
    Let Assumption~\ref{as:A-B-stabilizable} hold, then $\gamma^2 + \beta^2 > 0$.
    Furthermore, suppose that the conditions in Proposition~\ref{prop:cbf} hold. Then, $r^2 (\gamma^2 + \beta^2) - T_3^2 > 0$.
    Moreover, if Assumption \ref{as: interior eq}
    %in Lemma~\ref{lem:interior} 
    holds, then $\gamma x_{c,1} + \beta x_{c,2} \neq 0$.
\end{lemma}

\vspace{.1cm}

Next we characterize the undesirable equilibria of the closed-loop system~\eqref{eq:v-linear-system} with~\eqref{eq:linear-2d-underactuated-system}.
%when the system is of the form~\eqref{eq:linear-2d-underactuated-system}.

\vspace{.1cm}

\begin{proposition}\longthmtitle{Equilibria in Under-actuated Systems}\label{prop:undesired-eq-n-2}
    Suppose that Assumptions \ref{as: interior eq}, \ref{as:A-B-stabilizable} and the conditions in Proposition~\ref{prop:cbf} hold. Define  $\bp_{+} := ( \gamma z_{+}, \beta z_{+} )$, and $\bp_{-} := ( \gamma z_{-}, \beta z_{-} )$, where   \begin{align*}
        &z_{\pm} = \frac{ \gamma x_{c,1} + \beta x_{c,2} \pm \sqrt{ r^2 (\gamma^2 + \beta^2) - T_3^2 } }{ \gamma^2 + \beta^2 }.
    \end{align*}
    % \begin{align*}
    %     &\bp_{+} = ( \gamma z_{+}, \beta z_{+} ), \quad \bp_{-} = ( \gamma z_{-}, \beta z_{-} ) .
    % \end{align*}
    Then,
    \begin{enumerate}[i)]
        \item if $\gamma x_{c,1} + \beta x_{c,2} < 0$, $\bp_{+}$ is the only undesirable equilibrium of the closed-loop system~\eqref{eq:v-linear-system} with~\eqref{eq:linear-2d-underactuated-system};
        \item if $\gamma x_{c,1} + \beta x_{c,2} > 0$, $\bp_{-}$ is the only undesirable equilibrium of the closed-loop system~\eqref{eq:v-linear-system} with~\eqref{eq:linear-2d-underactuated-system}.
    \end{enumerate}
\end{proposition}

\vspace{.1cm}

Note that by Lemma~\ref{lem:discriminant-is-positive-if-cbf}, $\gamma x_{c,1} + \beta x_{c,2} \neq 0$. Therefore Proposition~\ref{prop:undesired-eq-n-2} shows that for linear, planar, underactuated and stabilizable linear systems,~\eqref{eq:general-system-1} has exactly one undesirable equilibrium.
Note also that the result in Proposition~\ref{prop:undesired-eq-n-2} is independent of the linear stabilizing controller $k$ and the extended class $\Kc$ function $\alpha$ chosen.

The following result establishes that the undesirable equilibrium of the closed-loop system is always a saddle point.

\vspace{.1cm}

\begin{proposition}\longthmtitle{Undesirable Equilibria are Saddle Points}\label{prop:undesired-eq-are-saddle-points}
    Suppose that Assumptions \ref{as: interior eq},  \ref{as:A-B-stabilizable} and the conditions in Proposition~\ref{prop:cbf} hold. Then there always exists one and only one undesirable equilibrium, which is a saddle point.
    
    %Then, if $\bp_{+}$ is an undesired equilibrium point of the closed-loop system, it is a saddle point.
    %Similarly, if $\bp_{-}$ is an undesired equilibrium point of the closed-loop system, it is a saddle point.
\end{proposition}

\vspace{.1cm}

Note that the results in Propositions~\ref{prop:undesired-eq-n-2} and~\ref{prop:undesired-eq-are-saddle-points} are independent of the choice of weighting matrix $G$, nominal controller $k$ or extended class $\Kc$ function $\alpha$. The combination of Propositions~\ref{prop:undesired-eq-n-2} and~\ref{prop:undesired-eq-are-saddle-points} with Theorem~\ref{thm:complete-characterization-planar-linear-underactuated} provide a complete picture of the under-actuated case, which we summarize as follows.

% \blue{
% \begin{corollary}
%      Suppose that Assumptions~\ref{as: interior eq},~\ref{as:A-B-stabilizable} and the conditions in Proposition~\ref{prop:cbf} hold.  Then for any Hurwitz $A-BK$, there exists a set $\mathcal{S}$ of measure $0$ and $\alpha_2^*>0$, such that  for any $\alpha(s)=\alpha_0 s$ with $\alpha_0\geq \alpha_2^*$ and any $G(\bx)$, the solution of \eqref{eq:v-linear-system} satisfies  $\lim\limits_{t\to\infty} \bx(t;\bx_0) = \textbf{0}_2$, for all $\bx_0\in\mathcal{C}\setminus \mathcal{S}$.
% \end{corollary}
% }
% \red{YC: I think Corallary \ref{cor:trajs-underactuated-2d-linear} is a bit repetitive compared to theorem \ref{thm:complete-characterization-planar-linear-underactuated}. I have written an alternative version above. We can also change the conclusion as the existence of unique undesirable eq and saddle point. It depends on what we want to emphasize.  } 
% \vspace{.1cm}

\begin{corollary}\longthmtitle{Characterization of trajectories for linear planar underactuated systems}\label{cor:trajs-underactuated-2d-linear}
    Suppose that Assumptions~\ref{as: interior eq},~\ref{as:A-B-stabilizable} and the conditions in Proposition~\ref{prop:cbf} hold. Then, the closed-loop system~\eqref{eq:v-linear-system} obtained from~\eqref{eq:linear-2d-underactuated-system} has one and only one undesirable equilibrium $\hat{\bx}^*$ equal to either $\bp_{+}$ or $\bp_{-}$.
    Additionally, there exists $\alpha_2^*>0$ such that for any $\alpha(s)=\alpha_0 s$ with $\alpha_0\geq \alpha_2^*$, if $W_s(\hat{\bx}^*)$ denotes the global stable manifold of $\hat{\bx}^*$ it holds that:
    \begin{enumerate}
        \item\label{it:first} if $\bx_0 \in W_s(\hat{\bx}^*)$, then $\lim\limits_{t\to\infty} \bx(t;\bx_0) = \hat{\bx}^*$; 
        \item\label{it:second} if $\bx_0 \notin W_s(\hat{\bx}^*)$, then $\lim\limits_{t\to\infty} \bx(t;\bx_0) = \textbf{0}_2$. \hfill $\Box$ 
    \end{enumerate}
\end{corollary}

\subsection{Fully Actuated LTI Planar Systems}

% \red{YC: We can try to include the proof for proposition \ref{Thm: number of equilibrium }, \ref{thm: m=2, x0 is not eigen, case 1}, \ref{thm: m=2, x0 is not eigen, case 2} if the space allows. As for the lemma \ref{lemma: compute eigen for m=2, case 1}, \ref{lemma: compute eigen for m=2, case 2} and tables, we can just state the result. The proof of lemma \ref{lemma: compute eigen for m=2, case 1} and \ref{lemma: compute eigen for m=2, case 2} is simple computation and the proof for the results in the table is too long.}

% \red{YC: not sure if we need to include proposition \ref{thm: m=2, x0 is not eigen, case 2} in this paper. It is correct but useless. }

We now consider the case where $B$ is invertible; in this case, Assumptions \ref{as: feasibility} and \ref{as:A-B-stabilizable} are satisfied.

\subsubsection{$\bx_c$ is an eigenvector of $\Tilde{A}$}  We start by considering two conditions for the case where $\bx_c$ is an eigenvector of $\Tilde{A}$.

\textit{Condition 1.} $\lambda_1<\lambda_2<0$, $\tilde{A}\bx_c=\lambda_2 \bx_c$, $\tilde{A} \bv_1=\lambda_1 \bv_1$, 
$\bv_2=\frac{\bx_c}{\|\bx_c\|}$,   $   1-\frac{(\lambda_1-\lambda_2)^2 r^2}{\lambda_2^2\|\bx_c\|^2}=0$, $(\bv_1^\top \bv_2)^2=   1-\frac{(\lambda_1-\lambda_2)^2 r^2}{\lambda_2^2\|\bx_c\|^2}$.

\textit{Condition 2.} $\lambda_1<\lambda_2<0$, $\tilde{A}\bx_c=\lambda_2 \bx_c$, $\tilde{A} \bv_1=\lambda_1 \bv_1$, $\bv_2=\frac{\bx_c}{\|\bx_c\|}$, $   1-\frac{(\lambda_1-\lambda_2)^2 r^2}{\lambda_2^2\|\bx_c\|^2}=0$, $(\bv_1^\top \bv_2)^2>   1-\frac{(\lambda_1-\lambda_2)^2 r^2}{\lambda_2^2\|\bx_c\|^2}$.

We have that there exists only one undesirable equilibrium and it is a degenerate equilibrium if and only if \textit{Condition 1} is true. 
%(the proof is omitted due to space limits).
If \textit{Condition 2} is true, there are two undesirable equilibria, one of which is a saddle point and the other one is a degenerate equilibrium.

If neither \textit{Condition 1} nor \textit{Condition 2} is true, we summarize the results about undesirable equilibria for the case that $\bx_c$ is an eigenvector of $\Tilde{A}$ in Tables \ref{tab:case_nondiagonalizable} and \ref{tab:case_diagonalizable}. We gather all the cases in the following result.
%
%\marginJC{This wording confused me a bit: the table summarize the result of the corollary? And this is a corollary of what result?}
%\marginYC{The tables summarize the results for full characterization of undesirable equilibria in the case that $\bx_c$ is an eigenvector of $\Tilde{A}$. The corollary 2 is the corollary of the two conditions and the tables. I have changed the Corollary to Proposition. }
%
% these tables  correspond to the case where $\tilde{A}$ is non-diagonalizable and the case where $\tilde{A}$ is diagonalizable, respectively.

\vspace{.1cm}

\begin{proposition}
\longthmtitle{Characterization of undesirable equilibria}\label{thm: m=2, x0 is  eigen}
    Let Assumptions \ref{as: interior eq} be satisfied and $B$ be invertible. Given that $\Tilde{A}$ is stable and $\bx_c$ is an eigenvector of $\tilde{A}$, then one of the following is true:
    \begin{itemize}
        \item[(i)] $|\mathcal{E}|=2$, $|\hat{\mathcal{E}}|=1$, $\bx\in  \hat{\mathcal{E}}$ is a degenerate equilibrium.

         \item[(ii)] $|\mathcal{E}|=2$, $|\hat{\mathcal{E}}|=1$,  $\bx\in  \hat{\mathcal{E}}$ is a saddle point.

           \item[(iii)] $|\mathcal{E}|=3$, $|\hat{\mathcal{E}}|=2$, one point in $\hat{\mathcal{E}}$ is a saddle point and the  other point in $\hat{\mathcal{E}}$ is a degenerate equilibrium.

           \item[(iv)] $|\mathcal{E}|=4$, $|\hat{\mathcal{E}}|=3$, two points in $\hat{\mathcal{E}}$ are saddle points and  the other point in $\hat{\mathcal{E}}$ is asymptotically stable. \hfill $\Box$
    \end{itemize}
\end{proposition}

\vspace{.1cm}

Proposition~\ref{thm: m=2, x0 is  eigen} asserts that the number and the stability property of the undesirable equilibria are determined by the number of solutions of \eqref{eq: condition-eq}, if $\bx_c$ is an eigenvector of $\tilde{A}$.

\vspace{.1cm}
   
\begin{proposition}\longthmtitle{Spectrum of $\tilde{A}$ does not determine stability properties of undesirable equilibria}\label{corollary: ambiguity}
  Let Assumption \ref{as: interior eq} be satisfied and $B$ be invertible. Then for any given negative $\lambda_1$ and $\lambda_2$, there exists $K_1$ and $K_2$ in the set $\{K: \lambda_1,\lambda_2 = \textrm{spec}(A-BK) \}$,  such that there is an undesirable asymptotically stable equilibrium after applying the CBF filter with $u=-K_1 \bx$; and there is only one undesirable equilibrium and it is a saddle point after applying the CBF filter with $u=-K_2 \bx$. \hfill $\Box$
\end{proposition}
%\red{YC: not sure if I need to say something like almost globally stable for $u=-K_2 \bx$. }
\vspace{.1cm}

\begin{table}[t!]
    \centering
    \begin{tabular}{l|c|c|c}
     &  \text{SP}  & \text{DE} & \text{ASE} \\  \hline
     $(\bv_1^\top \bv_2)^2<   1-\frac{ r^2}{\lambda^2\|\bx_c^2\|} $  &  1 & 0  &0\\
     $(\bv_1^\top \bv_2)^2=   1-\frac{ r^2}{\lambda^2\|\bx_c^2\|}$ &  1 & 1  &0\\
      $ (\bv_1^\top \bv_2)^2>   1-\frac{ r^2}{\lambda^2\|\bx_c^2\|} $  &  2 & 0  &1\\
    \end{tabular}
    \caption{$\tilde{A}$  stable, $\tilde{A} \bv_2=\lambda \bv_2+ \bv_1$, $ \bv_1=\frac{\bx_c}{\|\bx_c\|}$, $\tilde{A}\bx_c=\lambda \bx_c$, $\|v_2\|=1$. SP: saddle point, DE: degenerate equilibrium, ASE: undesirable asymptotically stable equilibrium.}
    \vspace{-.2cm}
\label{tab:case_nondiagonalizable}
\end{table}

\begin{table}[t!]
\vspace{-0.2cm}
    \centering
    \begin{tabular}{l|c|c|c}
    & \text{SP}  & \text{DE} & \text{ASE} \\ \hline 
    $ (\bv_i^\top \bv_j)^2<   1-\frac{(\lambda_i-\lambda_j)^2 r^2}{\lambda_i^2\|\bx_c\|^2}$   &  1 & 0  &0\\
    $(\bv_i^\top \bv_j)^2=   1-\frac{(\lambda_i-\lambda_j)^2 r^2}{\lambda_i^2\|\bx_c\|^2}$ &  1 & 1  &0\\
      $ (\bv_i^\top \bv_j)^2>   1-\frac{(\lambda_i-\lambda_j)^2 r^2}{\lambda_i^2\|\bx_c\|^2}$   &  2 & 0  &1\\
    \end{tabular}
    \caption{$\tilde{A}$ stable,  $\tilde{A}\bx_c=\lambda_i \bx_c$, $ \bv_i=\frac{\bx_c}{\|\bx_c\|}$, $\tilde{A} \bv_j=\lambda_j \bv_j$,  $\|\bv_j\|=1$, $i,j=\{1,2\}$,  $\{\bv_i, \bv_j\}$  linearly independent.}
    \vspace{-.8cm}
    \label{tab:case_diagonalizable}
\end{table}

Note that one can characterize the global stability properties of the origin based on the eigenvalues of $A-BK$. However, based on Proposition \ref{corollary: ambiguity}, the eigenvalues of $A-BK$ do not fully determine the global stability property of the origin. On the other hand, Proposition \ref{corollary: ambiguity} shows that there always exists a nominal controller $\bu=-K\bx$ such that $\tilde{A}$ has negative eigenvalues and the set of trajectories of~\eqref{eq:v-linear-system} that do not converge to the origin has measure zero (cf. Theorem~\ref{thm:complete-characterization-planar-linear-underactuated}). 
  Note that as shown in Lemma~\ref{lem:jacobian-characterization-general} and Tables \ref{tab:case_nondiagonalizable}, \ref{tab:case_diagonalizable}, the class $\Kc$ function only affects the rate of decay in the stable manifold of the undesirable equilibria and it does not affect the existence and stability of undesirable equilibria. Therefore, the choice of nominal controller $\bu=-K\bx$ determines in which of the cases we fall into. Ideally, the controller should be designed so that there exists only one undesirable equilibrium and it is a saddle point.

\subsubsection{$\bx_c$ is not an eigenvector of $\Tilde{A}$} Next, we analyze the number of undesirable equilibria when $\bx_c$ is not an eigenvector of $\tilde{A}$. In this case, the analysis is more involved and we only study the stability properties of undesirable equilibria under some sufficient conditions.

\vspace{.1cm}

\begin{proposition}[Number of undesirable equilibria]\label{Thm: number of equilibrium }
 Let Assumption \ref{as: interior eq} be satisfied and $B$ be invertible. Given that $G(\bx)=B^\top B$  and  $\tilde{A}$ is stable and  $\bx_c$ is not an eigenvector of $\Tilde{A}$, then  $ 1 \leq |\hat{\mathcal{E}}|\leq 3$ and $ |\mathcal{E}\setminus \hat{\mathcal{E}}|\geq 1$.
%and  $|\mathcal{E}|-|\hat{\mathcal{E}}|\geq 1$. 
 In addition, if $\lambda_1\leq \lambda_2$, there exists $\bx\in\mathcal{\hat{E}}$ with indicator $\delta <\frac{\lambda_1}{2}$. \hfill $\Box$
\end{proposition}

\vspace{.1cm}

Combining Propositions \ref{prop:ellipsoidal-circular-same-properties},  \ref{prop:undesired-eq-n-2}, \ref{Thm: number of equilibrium } and Table \ref{tab:case_nondiagonalizable},  \ref{tab:case_diagonalizable},  it follows that applying the CBF filter to a LTI planar system (either under or fully actuated) with a linear stabilizing controller always introduces at least one undesirable equilibrium when the obstacle is ellipsoidal. By~\cite[Thm. 9.5]{GT:12-ams} and Lemma~\ref{lem:jacobian-characterization-general}, there exists at least one trajectory converging to the undesirable equilibrium.   This result
is consistent with~\cite{PB-CMK:20}, which states that given a local Lipschitz dynamical system and a compact unsafe set, if the safe set is forward invariant then there exists at least one trajectory that does not converge to the origin.  Theorem \ref{thm:complete-characterization-planar-linear-underactuated} ensures that if there is only one undesirable equilibrium and it is a saddle point, then there is only one such trajectory and it corresponds to the global stable manifold of the undesirable equilibrium.

To analyze the stability of undesirable equilibria in the case that  $\bx_c$ is not an eigenvector of $\tilde{A}$, we need to determine the eigenvalues of $J\mid_{\bx\in\hat{\mathcal{E}}}$. By Lemma \ref{lem:jacobian-characterization-general}, $-\alpha^\prime(0)=-\alpha_0$ is an eigenvalue of $J\mid_{\bx\in\hat{\mathcal{E}}}$. 
{ The result in~\cite[Lemma 5]{extended} provides an expression for the other eigenvalue of $J\mid_{\bx\in\hat{\mathcal{E}}}$, and by leveraging it, we get the following result.}

\begin{figure*}[h!]
  \centering 
  {\includegraphics[width=0.9\textwidth]{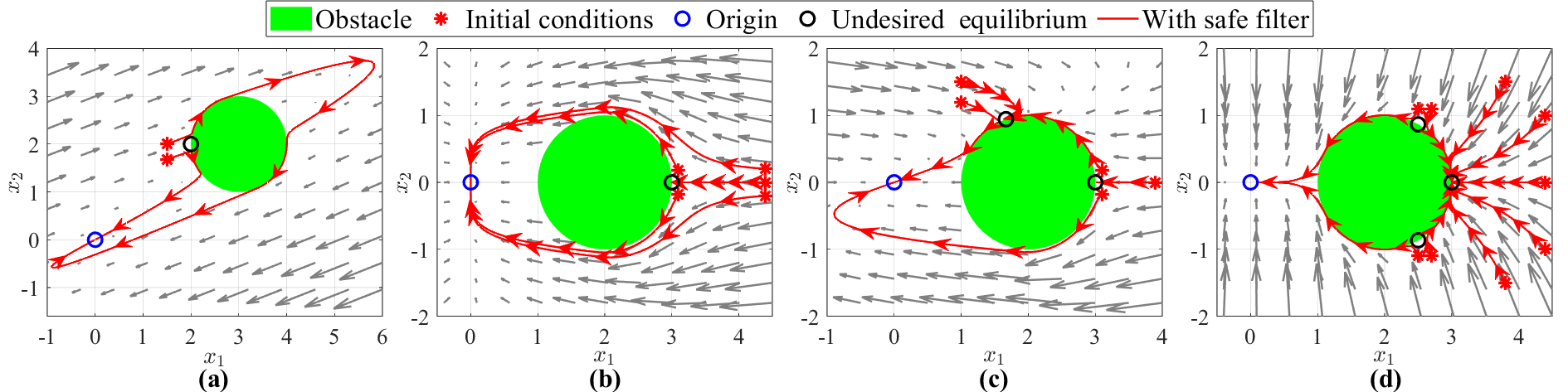}} 
   \vspace{-.2cm}
  \caption{Examples of
trajectories of an LTI planar system with a safety filter for a circular obstacle; the figures show the vector fields, the undesirable equilibria, and the desired  equilibrium (which is the origin). (a): Under-actuated system. (b)-(c)-(d): Fully actuated system,  corresponding to the three rows of Table \ref{tab:case_diagonalizable} respectively. In (a) and (b) the undesirable equilibrium is a saddle point. In (c) there is one degenerate equilibrium and one saddle point. In (d) there are three undesirable equilibria,  one is asymptotically stable while the others are saddle points. }
  \label{fig: under-actuated case and fully actuated cases}
   \vspace{-.5cm}
\end{figure*}

\vspace{.1cm}

\begin{proposition}\longthmtitle{Sufficient conditions for undesirable equilibria}\label{thm: m=2, x0 is not eigen, case 1}
 Let Assumption \ref{as: interior eq} be satisfied and $B$ be invertible. Given that $G(\bx)=B^\top B$,  $\tilde{A}$ is stable  with two real eigenvalues $\lambda_1<\lambda_2$ and  $\bx_c$ is not an eigenvector of $\Tilde{A}$, then there is no undesirable equilibrium with indicator $\delta\in\{\frac{\lambda_1}{2}, \frac{\lambda_2}{2}\}$ .
 Besides, let $\bv_1$ and $\bv_2$ be the eigenvectors associated with $\lambda_1$ and $\lambda_2$, respectively, and $\bv_1^\top \bv_2\geq 0$, $\|\bv_1\|=\|\bv_2\|=1$; and then  we can write $\bx_c=\beta_1 \bv_1+\beta_2 \bv_2$. Then, the following holds.
 \begin{itemize}
     \item[i)] If $\beta_1^2+\beta_1 \beta_2 \bv_1^\top \bv_2\geq 0$, then for any undesirable equilibrium $\bx$ with indicator $\delta$ such that $\delta<\frac{\lambda_1}{2}$,  $\bx$ is a saddle point.

  \item[ii)] If $\beta_1\beta_2
 \bv_2^\top \bv_1+\beta_2^2\geq 0$, then for any undesirable equilibrium $\bx$ with indicator $\delta$ such that $\frac{\lambda_2}{2}<\delta<0$,  $\bx$ is asymptotically stable.
     
     \item[iii)] Define $F_1:\mathbb{R}\to \mathbb{R}$ as:
\begin{equation}\label{eq: F_1}
\begin{aligned} 
    F_1(\delta)&:=-|\lambda_1-2\delta|^2 |\lambda_2-2\delta|^2 r^2+|\lambda_1\beta_1|^2 |\lambda_2-2\delta|^2\\
    &+|\lambda_2\beta_2|^2 |\lambda_1-2\delta|^2\\
    &+2\text{Re}(\lambda_1^*\beta_1^*\lambda_2\beta_2(\lambda_2-2\delta)^*(\lambda_1-2\delta)  \bv_1^* \bv_2).
\end{aligned}
\end{equation}
If the third order polynomial $\frac{d F_1(\delta)}{d \delta}$  has only one real root\footnote{
    For third-order polynomial $ax^3+bx^2+cx+d$, its discriminant is defined as $18 a b c d-4 b^3 d+b^2 c^2-4 a c^3-27 a^2 d^2$. If $a\neq 0$ and the discriminant is negative, the third-order polynomial only has one real root. } and $\beta_1^2+\beta_1 \beta_2 \bv_1^\top \bv_2\geq 0$, then there exists only one undesirable equilibrium and it is a saddle point. \hfill $\Box$
 \end{itemize}
\end{proposition}

\vspace{.1cm}

If $|\beta_1|\gg|\beta_2|$ and $|\bv_1^\top \bv_2|$ is small, then the case $\beta_1^2+\beta_1 \beta_2 \bv_1^\top \bv_2\geq 0$ is a generalized version of the case in the first row of Table \ref{tab:case_diagonalizable}. If $|\beta_2|\gg|\beta_1|$ (i.e., $\bx_c$ is ``essentially'' eigenvector associated with $\lambda_2$) and $\lambda_1\ll \lambda_2$, then the case $\beta_1\beta_2
 \bv_2^\top \bv_1+\beta_2^2\geq 0$ is a generalized version of the case in the last row of Table \ref{tab:case_diagonalizable}, as  $1-\frac{(\lambda_2-\lambda_1)^2 r^2}{\lambda_2^2\|\bx_c\|^2}<0$ with $\lambda_1\ll\lambda_2$.

\section{Numerical Experiments}
%
% \marginJC{In this section, as we go to each of the examples and show what happens, it would be nice to link back to the results earlier: e.g., say things like "as guaranteed by Proposition~BLA" or "consistently with the result in Proposition~BLA"}
%
As a first experiment, we consider the safety set $\mathcal{C} = \{\bx:~\|\bx-(3,2)^\top\|^2-1\geq 0 \}$ and the under-actuated system $\dot \bx=\begin{bmatrix}
    4& 2 \\  1 &   1
\end{bmatrix}\bx+\begin{bmatrix}
    3\\1
\end{bmatrix}\bu$ with nominal controller $\bu=-\begin{bmatrix}
    3 & -2
\end{bmatrix}\bx$. Once the CBF-based filter is applied, there is one undesirable equilibrium $(2,2)^\top$, as guaranteed by Proposition~\ref{prop:undesired-eq-n-2}. Examples of trajectories of the system with the safety filter are shown in Figure \ref{fig: under-actuated case and fully actuated cases}(a), along with the vector field, the spurious undesirable equilibrium, and the desirable equilibrium (which is the origin).

In Figures \ref{fig: under-actuated case and fully actuated cases}(b), (c) and (d), we consider a safety set $\mathcal{C} = \{\bx:~\|\bx-(2,0)^\top\|^2-1\geq 0 \}$, and the integrator dynamics $\dot \bx=\bu$ as an example of \eqref{eq:2d-linear-system}. 

In Figure \ref{fig: under-actuated case and fully actuated cases}(b), we show the results for the integrator dynamics with $K=\begin{bmatrix}
-5 & 0\\ 0& -1
\end{bmatrix}$, $G(\bx)=B^\top B$, and  the safety filter with $\alpha(s)=\alpha_0 s$, $\alpha_0=10$. There is one undesirable equilibrium $(3,0)^\top$. We note that for both the setups in Figures \ref{fig: under-actuated case and fully actuated cases} (a) and (b), there is only one undesirable equilibrium and it is a saddle point. Only one trajectory converges to the undesirable equilibrium and all other trajectories converge to the origin.

In Figure \ref{fig: under-actuated case and fully actuated cases}(c), we show the results for \eqref{eq:v-linear-system} with $K=\begin{bmatrix}
-3 & 4\sqrt{2}\\ 0& -1
\end{bmatrix}$, $G(\bx)=B^\top B$ and $\alpha(s)=\alpha_0 s$, $\alpha_0=10$. There two undesirable equilibria, which are $(\frac{5}{3},\frac{2\sqrt{2}}{3})^\top$ (degenerate equilibrium) and $(3,0)^\top$ (saddle point). Only one trajectory converges to $(3,0)^\top$. The measure of the stable set of the degenerate equilibrium is positive (in fact, the measure is $+\infty$), although the degenerate equilibrium is unstable. 

In Figure \ref{fig: under-actuated case and fully actuated cases}(d),  we show that results for \eqref{eq:v-linear-system} with $K=\begin{bmatrix}
-1 & 0 \\ 0& -5
\end{bmatrix}$,  $G(\bx)=B^\top B$ and $\alpha(s)=\alpha_0 s$, $\alpha_0=10$. There are three undesirable equilibria: $(\frac{5}{2},\frac{\sqrt{3}}{2})^\top$, $(\frac{5}{2},-\frac{\sqrt{3}}{2})^\top$ and $(3,0)^\top$; the last one is asymptotically stable and the first two are saddle points. The two trajectories converging to  $(\frac{5}{2},\frac{\sqrt{3}}{2})^\top$, $(\frac{5}{2},-\frac{\sqrt{3}}{2})^\top$ and part of the obstacle constitute the boundary of the region of attraction of $(3,0)^\top$. Since the examples in Figure~\ref{fig: under-actuated case and fully actuated cases}(b), (c) and (d) all satisfy that $\bx_c$ is an eigenvector of $\tilde{A}$, these results are consistent with Proposition~\ref{thm: m=2, x0 is  eigen} (ii), (iii), (iv), respectively.

% % %% test
% \begin{figure*}[h!]
%     \centering
%     \subfigure[]{\includegraphics[width=0.32\textwidth]{Figures/cbf_n2_m1_traj_vector_field.png}}
%     \subfigure[]{\includegraphics[width=0.32\textwidth]{Figures/cbf_m2_n2_one_eq_traj_vector_field.png}}
%     \subfigure[]{\includegraphics[width=0.32\textwidth]{Figures/cbf_n2_m1_traj_vector_field.png}}
%     \caption{(a) blah (b) blah (c) blah}
%     \label{fig:foobar}
% \end{figure*}
% % %%

% \begin{figure}[h!]
%   \centering 
%   {\includegraphics[width=0.46\textwidth]{Figures/cbf_n2_m1_traj_vector_field.png}} 
%   \vspace{-.6cm}
%   \caption{$m=1$}
%   \label{fig: m=1, 1 eq}
% \end{figure}

% \begin{figure}[h!]
%   \centering 
% {\includegraphics[width=0.46\textwidth]{Figures/cbf_m2_n2_two_eq_traj_vector_field.png}} 
% \vspace{-.4cm}
%   \caption{$m=2$, one equilibrium}
%   \label{fig: m=2, 1 eq}
% \end{figure}

% \begin{figure}[h!]
%   \centering 
% {\includegraphics[width=0.46\textwidth]{Figures/cbf_m2_n2_two_eq_traj_vector_field.png}} 
% \vspace{-.7cm}
%   \caption{$m=2$, two equilibria}
%   \label{fig: m=2, 2 eq}
% \end{figure}

% \begin{figure}[h!]
%   \centering 
%  {\includegraphics[width=0.46\textwidth]{Figures/cbf_m2_n2_three_eq_traj_vector_field.png}} 
% \vspace{-.4cm}
%   \caption{$m=2$, three equilibria}
%   \label{fig: m=2, 3 eq}
%   \vspace{-.7cm}
% \end{figure}

\bibliographystyle{IEEEtran}
\bibliography{alias,Main,Main-add,references}

\end{document}